\documentclass[a4paper, amsfonts, amssymb, amsmath, reprint, showkeys, nofootinbib, twoside, superscriptaddress,longbibliography]{revtex4-2}
\usepackage[dvipsnames]{xcolor}
\usepackage[english]{babel}
\usepackage[utf8]{inputenc}
\usepackage[colorinlistoftodos, color=green!40, prependcaption]{todonotes}
\usepackage{amsthm}
\usepackage{mathtools}
\usepackage{physics}
\usepackage{xcolor}
\usepackage{graphicx}
\usepackage[left=23mm,right=13mm,top=35mm,columnsep=15pt]{geometry} 
\usepackage{adjustbox}
\usepackage{placeins}
\usepackage[T1]{fontenc}
\usepackage{lipsum}
\usepackage{csquotes}
\usepackage[pdftex, pdftitle={Article}, pdfauthor={Author}]{hyperref} 
\usepackage{upgreek}
\usepackage[normalem]{ulem}
\usepackage{float}
\begin{document}
\title{Effects of chlorine substitution on the magneto-optical properties of CrSBr}

\newcommand{\affiliationUW}{\affiliation{Faculty of Physics, University of Warsaw, Warsaw, 02-093 Poland}}
\newcommand{\affiliationCzech}{\affiliation{Department of Inorganic Chemistry, University of Chemistry and Technology Prague, Technicka 5, 166 28 Prague
6, Czech Republic}}

\author{Z. Śnioch}
\affiliationUW

\author{R. Komar}
\affiliationUW

\author{M. Kuna}
\affiliationUW

\author{M. Raczyński}
\affiliationUW

\author{ K. Mosina}
\affiliationCzech

\author{A.~S\"{o}ll}
\affiliationCzech

\author{Z. Sofer}
\affiliationCzech

\author{K. Wohlfeld}
\affiliationUW

\author{P. Kossacki}
\affiliationUW

\author{T. Kazimierczuk }
\affiliationUW


\begin{abstract}
We present a systematic study of the effects of chlorine substitution on the magneto-optical properties of the van der Waals layered semiconductor CrSBr. Our study shows that even at substitution level of $x=56\%$ the $\mathrm{CrSBr_{1-x}Cl_{x}}$  material exhibits qualitatively the same clear magneto-optical coupling that was observed in pure CrSBr crystals. The values of the relevant parameters are shown to vary smoothly across different substitution levels, thus offering promising tunability for this family of 2D magnetic materials.

\end{abstract}

\keywords{two-dimensional materials, magnetic semiconductor materials, halogen substitution}

\maketitle

\section{Introduction} \label{sec:outline}
Chromium sulfur bromide (CrSBr) is a van der Waals layered
semiconductor which exhibits a magnetic ordering at cryogenic
conditions~\cite{crsbr-magnetoresistance-2020}. The remarkable
characteristic of CrSBr is an exceptionally strong coupling between
its magnetic and excitonic properties~\cite{crsbr-wilson-2021}. This
interplay allows direct probing of the magnetic state by means of
optical spectroscopy, unlocking observation of a vast variety of
magnetic phenomena in low-dimensional system. A number of studies
utilized this optical access to study, e.g., strain tuning of the
magnetism of thin flakes of CrSBr~\cite{cenker_reversible_2022} or
propagation of the photo-excited
magnons~\cite{bae_exciton-coupled_2022}. Unfortunately, CrSBr is
unique in a sense that even chemically close crystals such as
CrSeBr~\cite{han-crsebr} or CrOCl~\cite{zhang-crocl} were not found to offer
such direct link between their magnetization state and the exciton
energy, hindering development of the CrSBr-based heterostructures.
However, this gap in the material availability can be bridged by a
partial ligand substitution, such as replacing some Br atoms with Cl
or I. Feasibility of such approach was already demonstrated in
magnetometry measurements supported by the DFT modeling~
\cite{telford_designing_2023}. Here we further explore the effect of
Cl substitution, with a particular focus on the core optical and
magnetic properties of the mixed crystals.
During the preparation of this manuscript, a study on similar crystals
by S. Badola et al. was published~\cite{badolaVanWaalsCrSBr2026}.
Although the scope of the two studies differs, the results in the
overlapping regime are consistent and provide an independent point of
comparison.

\begin{figure}
    \centering
    \includegraphics{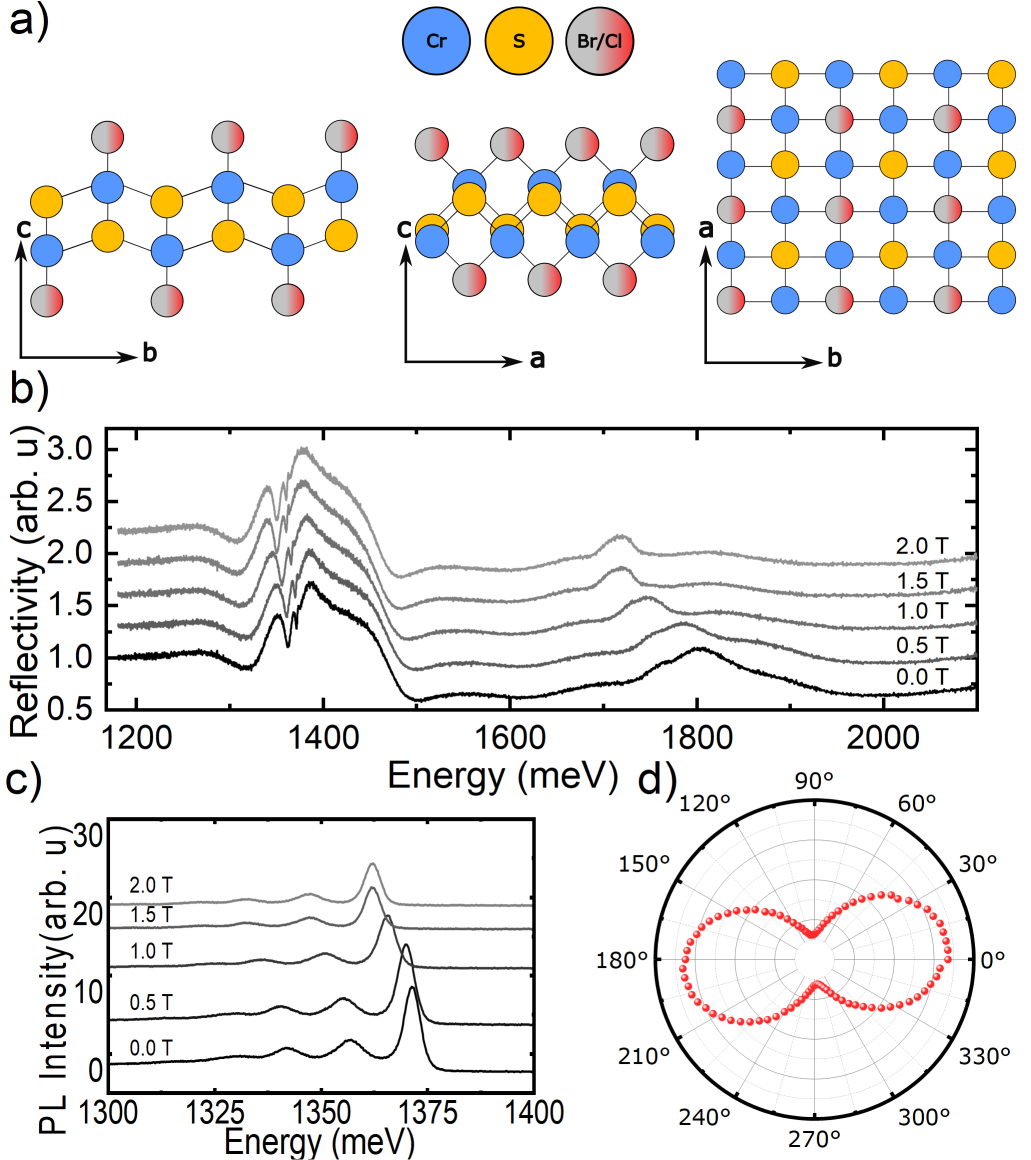}
    \caption{a) Crystal structure of CrSBr mixed crystals as viewed along the b-axis, a-axis, and c-axis  Cr, S, Br, Cl/I atoms are false colored to appear blue, yellow, red and grey, respectively b) reflectivity spectrum of CrSBr$_{0.56}$Cl$_{0.44}$ measured in differend magnetic fields in \textit{b} polarization. c) photoluminescence spectrum of CrSBr$_{0.56}$Cl$_{0.44}$ measured in differend magnetic fields in \textit{b} polarization. d) anisotropy of the fitted absorption state intensities as a function of light
polarization angle  for excitonic state $\Omega_1$.}
    \label{fig:1}
\end{figure}

We discuss our data with respect to the known properties of the
pristine CrSBr and CrSCl crystals. Structurally, CrSBr has an
orthorombic crystal structure~\cite{goser_magnetic_1990}. It consists
of covalently bonded CrS planes terminated with bromine atoms and
stacked via weak van der Waals forces along the c-axis
Fig.~\ref{fig:1} a). It exhibits A-type antiferromagnetic order:
within each layer, spins are aligned ferromagnetically along the
b-axis, while adjacent layers couple
antiferromagnetically~\cite{crsbr-wilson-2021}. This spin arrangement
remains stable up to the Néel temperature of approximately 132
K~\cite{lee_magnetic_2021}, providing an extended temperature window
for exploring magneto-optical phenomena in CrSBr compared to many
other magnetic semiconductors.
Substituting the outer bromine (Br) ligands in CrSBr with chlorine substantially
changes both the crystal structure and the magnetic properties of the
material. The strongly anisotropic in-plane rectangular lattice of
CrSBr is transformed into a triangular plane for pure
CrSCl~\cite{wangPTypeVdWSemiconductor2025}. In consequence, the
geometry of the interlayer super-exchange changes and the magnetic
coupling between layers switches the character from antiferromagnetic
to ferromagnetic~\cite{wangPTypeVdWSemiconductor2025}, what can be
understood in the frame of Goodenough-Kanamori-Anderson (GKA)
rules~\cite{GKA1, GKA2} .

Such fundamental changes in both structural and magnetic properties between CrSBr and CrSCl hint that the transition between these edge cases does not simply follow a virtual crystal approximation model. By a gradual Cl substitution we can explore the evolution of the material properties and uncover the gradual switching from anisotropic to isotropic in-plane system.
These properties can be accessed optically, by probing various excitonic transitions hosted by the alloys in static and, more robust, dynamic spin conditions.

\section{Samples  and methods} \label{sec:develop}

Our study is based on a systematic series of five $\mathrm{CrSBr_{1-x}Cl_x}$ crystals with \textit{x} ranging from 11\% to 56\%, as well as the pristine CrSBr. All the crystals were synthesized using chemical vapor transport (CVT)  using elemental chromium, sulfur, bromine and chromium chloride as  starting materials. The samples were prepared by mechanical exfoliation onto silicon substrates with a 90 nm SiO${_2}$ layer, chosen to optimize optical contrast in the excitonic range. Flake thickness was verified using atomic force microscopy (AFM), yielding approximately 20 nm thickness, equivalent to $\sim{30}$ layers. 
With increasing Cl content the crystal morphology noticeably evolved: the flakes become narrower and more needle-like; yet, this tendency did not hinder the feasibility of obtaining a~uniform flakes large enough for the subsequent optical experiments.

The core results of our work were obtained by means of magneto-optical measurements, which were performed at $T = 5$ K in a helium-cooled cryostat. Magnetic field was applied in the Faraday geometry, along the crystallographic \textit{c}-axis. The sample was mounted on a three-axis piezoelectric stage, enabling precise positioning and focus control. The optical measurements were carried out using free-beam optics. Focusing was accomplished by utilizing a $f=7.5$-mm achromatic lens, to a spot of a diameter of  $\approx 2$-$\mu$m. 
In case of the reflectivity measurements, the sample was illuminated using broadband white light from the tungsten light source. The PL measurements were performed using 635 nm laser excitation. In both cases the optical signal from the sample was detected using a Czerny–Turner spectrometer with a CCD camera.

As described in the following sections, in our work we investigated also the frequency of magnon excitation for different Cl content. These data were accessed in a further series of time-resolved pump-probe measurements, similar to Ref. \cite{bae_exciton-coupled_2022}. Here, the samples were excited using a strong femtosecond pulse at $\lambda=515$ nm and transient reflectivity was recorded using a second, time-synchronized femtosecond pulse in $\lambda=910$--$920$ nm range.
The time separation between the two pulses was adjusted using a mechanical delay line. The intensity of the reflected probe pulse exhibited clear oscillatory behavior in the GHz range, directly linked with the temporal evolution of the magnetization in the studied samples \cite{Diederich-Tunable-Interaction-Excitons-2023}.

\section{Effect of chlorine content on the optical properties \label{sec:develop}}  
In our study we employed several distinct experimental techniques to assess the optical properties: Raman scattering, reflectivity and photoluminescence (PL) measurements. In particular, the  Raman scattering evidences evolution of the optical phonon frequencies as a function of Cl- content, already showing that the studied crystals might be considered as a uniform material with the same qualitative features (e.g., in terms of observed vibration modes), yet differing quantitatively in their properties.
 However, as the Raman properties were already analyzed in-depth in a recent work by Badola \textit{et al.} \cite{badolaVanWaalsCrSBr2026}, our Raman-scattering results are presented in the Supplementary Information.

 Here, we focus on the PL and reflectivity experiments, which are pertinent for the time-resolved results presented in a further section. For each of the studied samples, the signal exhibited clear linear polarization dependence with principal axes corresponding to \textit{a}- and \textit{b}- crystallographic directions (
Fig. 1(d)). While the signal (reflectivity and PL) in the polarization \textit{a} consisted of only smooth background, without notable structure, the signal in the polarization \textit{b} exhibited distinct features attributed to the optically-active excitonic transitions. 

The photoluminescence of all the samples was predominantly in the near infrared region of 1.3 -- 1.4 eV (900 -- 950 nm). This emission corresponds to the ground state of the Frenkel-like exciton in CrSBr \cite{Datta-Magnon-mediated-ExcitonExciton-2025} (denoted here as $\Omega_1$). For the sake of simplicity, we estimated the energy of this exciton transition as the energy of the highest energy one of the several peaks in the spectrum, which was usually the strongest one (see Fig. 1c). Lower energy peaks were interpreted as either higher-order polariton-resonances \cite{wang-Exciton-polaritons2023} or localized states \cite{Smiertka-Distinct-Magnetooptical-Response2026}.

In the same energy range, the reflectivity spectra also featured sharp, clearly distinguished resonances. The additional benefit of the reflectivity approach was its sensitivity to higher-energy excitons known to exist in CrSBr around energy of 1.6 eV \cite{ Rafał-arxiv,Smiertka-Distinct-Magnetooptical-Response2026}, denoted here as $\Omega_3$ and $\Omega_4$ for the sake of  consistency with Ref. \onlinecite{Rafał-arxiv}.

Importantly, these qualitative features were observed for all the studied samples of CrSBr$_{1-x}$Cl$_{x}$. The extracted values, however, reveal a robust systematic trend as a function of halogen content: increasing chlorine content shifts the excitonic resonances to higher energies. 
As shown in Figure 2 a-c, this shift is observed for all identified excitonic states $\mathrm{\Omega_1}$--$\mathrm{\Omega_4}$. 
This effect is especially notable for the higher-energy $\mathrm{\Omega_3}$ exciton, which energy increases by nearly 40 meV between the pristine crystal and the 56\% Cl--substituted sample. This shift is about twice as large as that observed for lower-energy excitons. In contrast, the $\mathrm{\Omega_4}$ resonance, though close in energy to $\mathrm{\Omega_3}$, shows a weaker dependence on composition. This indicates that it couples less strongly to the electronic bands composed mainly from the halogen-ion orbitals.

The absolute energy of the excitonic states discussed above is a crucial factor determining bandgap alignment in heterostructures. Yet, in case of the CrSBr equally important is the issue of coupling between the excitons and the magnetic order. A typical behavior is shown in Fig. 1(b-c). Upon increasing external magnetic field along the \textit{c} axis, the excitonic resonances experience a red-shift. The underlying reason for the energy shift is that the magnetic field is gradually aligning the magnetic moments of the initially antiparallel-oriented sublattices (layers of the A-type antiferromagnet). The more magnetic order deviates from the strictly antiparallel configuration, the inter-layer hopping term for both carriers allow the excitons to delocalize more, lowering their energy \cite{crsbr-wilson-2021}. In the strong field limit, the adjacent layers become ferromagnetically oriented with no obstacle to exciton delocalization. In this limit the exciton energy remains practically constant, with exception of a residual effect of the diamagnetic shift \cite{Smiertka-Distinct-Magnetooptical-Response2026}.
More detailed discussion of the origin of the field-induced energy shift of the exciton can be found in Ref. \onlinecite{heissenbuttel-quadratic-response-crsbr}.

\begin{figure} 
    \centering
    \includegraphics[width =1\linewidth]{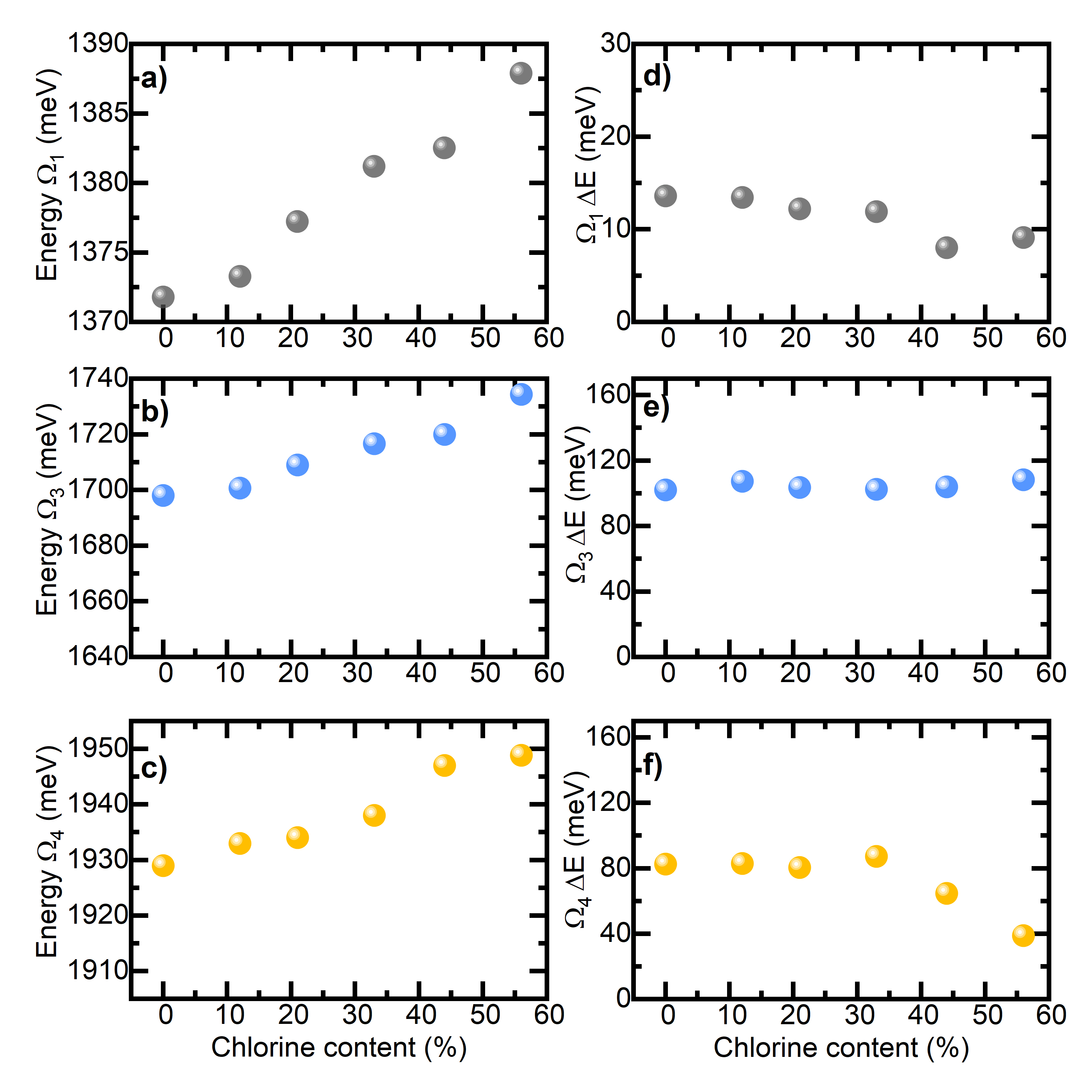}
    \caption{ Energy of each exciton as a function of the sample composition --  a, b, c.  Energy shift of each exciton as a function of the sample composition -- d, e, f.}
    \label{fig:3}
\end{figure}

Pertinent to the current study is the question, to what extent the chlorine substitution affects the discussed coupling between the excitons and the magnetic order. From the practical standpoint, the measure of this coupling is the energy difference between excitonic transition in the anti-parallel and parallel spin configurations, i.e., the state at $B=0$ T and at $B \gtrsim  2.5$ T. 
The experimental data extracted for each of the excitons identified in the spectrum are shown in Figs. \ref{fig:3} d,e,f. The lowest-energy exciton $\Omega_1$ was characterized using both PL and reflectivity, while higher energy excitons rely solely on the reflectivity measurements.

The extracted data evidences surprisingly weak dependence of the coupling strength. E.g., while 56\% Cl shifts the energy of $\Omega_3$ exciton up by about 40 meV, the corresponding differences between anti-parallel and parallel spin configurations are the same within precision of single meV. Such a property reveals intriguing prospects for future studies, such as  that the confinement in CrSBr$_{1-x}$Cl$_x$/CrSBr heterostructures would remain the same in both the low-field and high-field states, whereas the absolute exciton energy would still change significantly as known for pristine CrSBr.

The observed trends may be influenced by the effective ionic radius and electronegativity of the substituents.  In particular, differences in electronegativity among Cl and Br might alter local bonding and charge distribution, possibly providing an additional mechanism for tuning the excitonic spectrum. While the magnitude of this effect is exciton-dependent, overall, halogen substitution markedly modifies both the absolute exciton energies and their magnetic-field response in CrSBr-based crystals.\par

\section{Effect of Cl-substitution on the magnetic properties}  \label{sec:develop}
Halogen substitution affects not only the excitonic resonances but also the magnetic order itself. Although these data were extracted from the analysis of reflectivity and PL spectra, we emphasize that the exciton density in our experiments was sufficiently low to preserve the intrinsic behavior of the magnetic system.

Although optical measurements cannot deliver information on the absolute magnetization of the material, by tracing the excitonic shift as a function of the external magnetic field, we study the changes in the relative magnetization of the adjacent layers. In particular, upon increasing magnetic field along the \textit{c} axis, the excitons initially experience clear redshift (see Fig. 1b). Then, in a well-defined critical field $B_0$ the shift stops changing and further increase of the magnetic field does not affect exciton energy. This critical field $B_0$ corresponds to the point at which all the spins became fully aligned along \textit{c} directions.
As shown in Ref. \onlinecite{crsbr-wilson-2021}, the spin-canting behavior leads to magnetization given up to the saturation field  by $M=M_\mathrm{sat}\cos\left(\theta/2\right)\propto B_z$, where $\theta$ is the angle between magnetization vectors of the two sublattices.

The saturation field $B_0$ can be directly linked to the magnetic anisotropy parameter, and thus serves as a probe of the intrinsic magnetic properties of this material. This point will be discussed in detail later in this section. 

\begin{figure}
    \centering
    \includegraphics[width =1\linewidth]{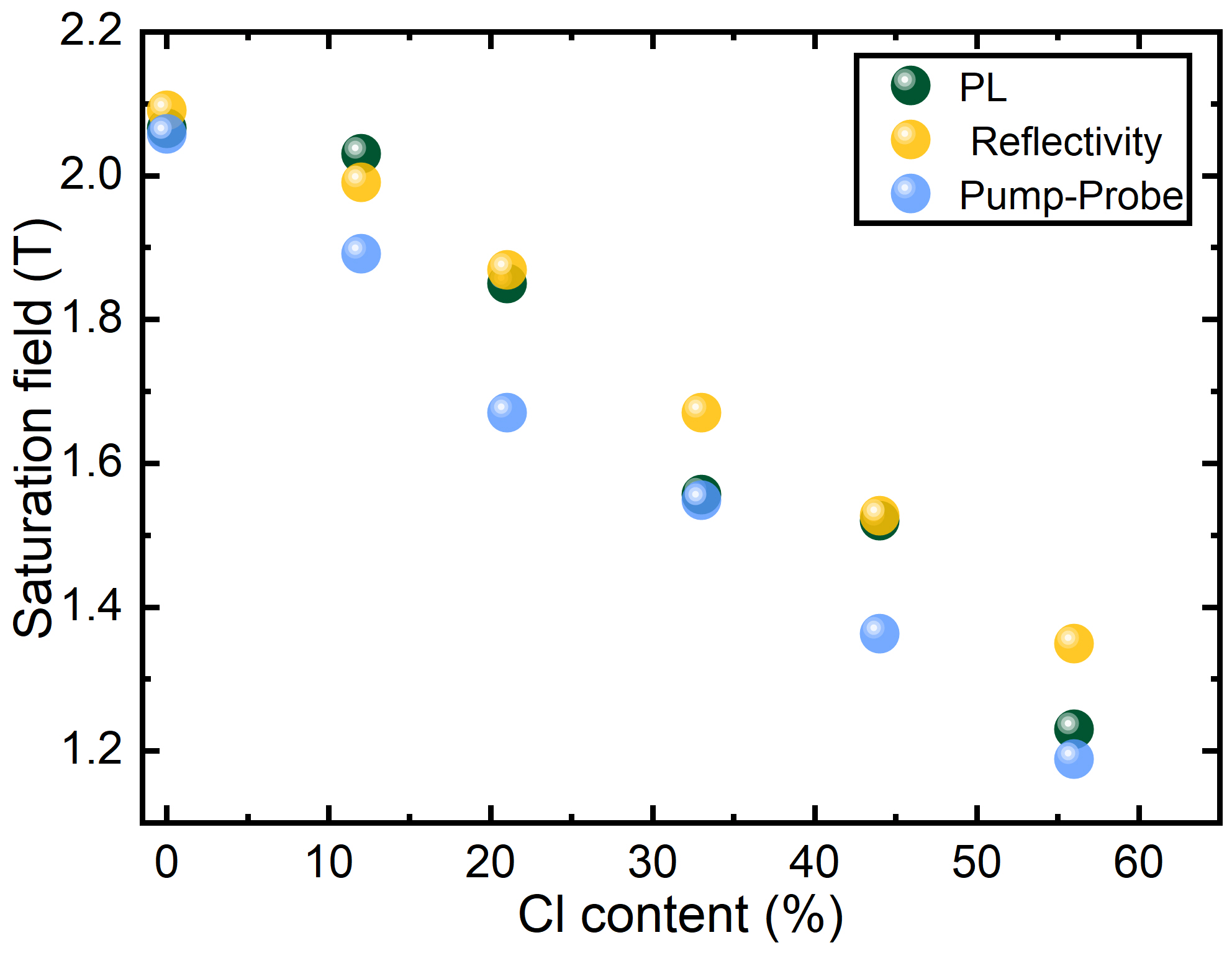}
    \caption{Saturation field as a function of sample composition, measured using reflectivity (yellow), photoluminescence (green) and pump-probe (blue) experiments.}
    \label{fig:4}
\end{figure}

In our experiments, the saturation field was extracted separately both from the reflectivity and the PL measurements. Both techniques reveal a consistent trend: increasing chlorine content results in a systematic decrease in the saturation field, down to the values approximately 0.7 T lower than those of the pristine crystal (Figure \ref{fig:4}). 
This behavior is explained by the role of halogen-mediated interlayer exchange interactions. In Cr–Br–Br–Cr superexchange chains, the extended 4p orbitals of bromine facilitate stronger antiferromagnetic coupling than the shorter 3p orbitals of chlorine in Cr–Cl–Cl–Cr configurations \cite{telford_designing_2023}. As a result, a lower external field is required to induce a ferromagnetic state in chlorine-substituted crystals. 


\begin{figure*}
    \centering
    \includegraphics{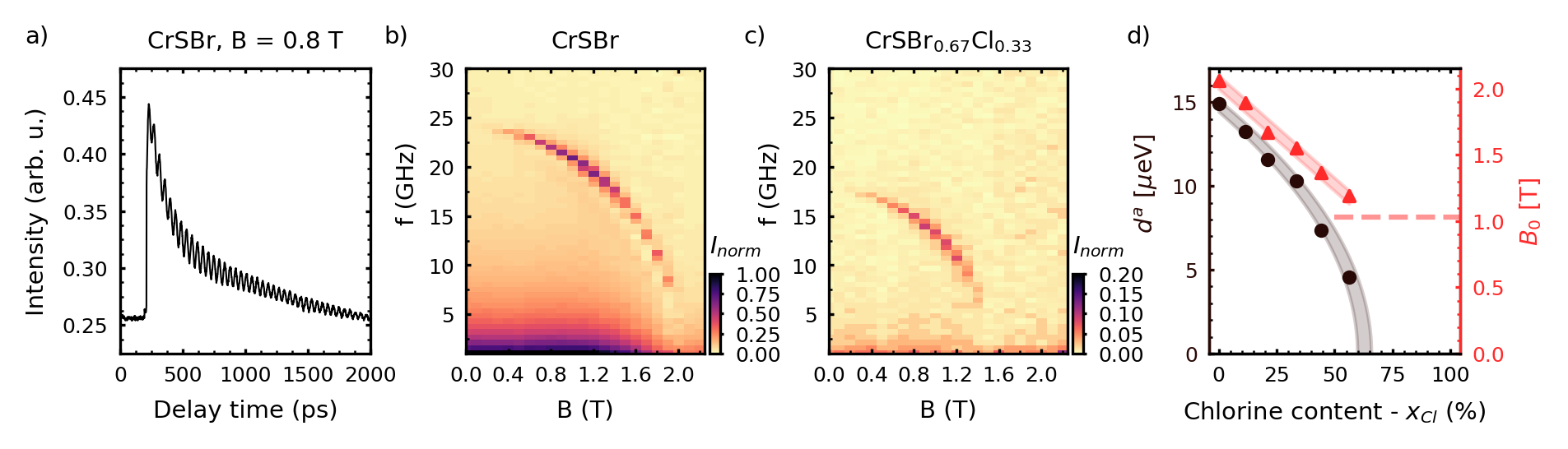}
    \caption{a) Transient optical reflectivity as a function of pump–probe delay at B = 0.8 T.
    b) Magnon dispersion map of CrSBr, obtained by Fourier transform of the data in a.
    c) Magnon dispersion map of CrSBr$_{0.67}$Cl$_{0.33}$. In figures b) and c) The intensity was normalized to the maximum in map b).
    d) Parameters fitted to the magnon dispersion maps: magnetic anisotropy in \textit{a}-axis -- $d^a$ and saturation field -- $B_0$ as a function of chlorine content. The shaded regions serve as a guide to the eye for trends of parameters' change. The dashed line marks the saturation field expected for pure CrSCl, based on parameters in Ref. \onlinecite{wangPTypeVdWSemiconductor2025}.}
    \label{fig:magnony}
\end{figure*}

As indicated above, the measurements in static spin conditions give us
insight into the intrinsic magnetic properties, such as the magnetic
anisotropy. Complementary data can be accessed using dynamical
conditions, corresponding to the controlled excitation of coherent
magnons \cite{bae_exciton-coupled_2022,Diederich-Tunable-Interaction-Excitons-2023}.
Performing the Linear Spin Wave Theory (LSWT, more detailed derivation
available in [SM]) on the magnetic Hamiltonian for rectangular-lattice
system with single-ion anisotropy fields and a magnetic field applied
along \textit{c}-axis, well describing CrSBr, we obtain two
eigenmodes, corresponding to the two magnon branches in the material.
In the experiments we are interested in the lower-energy excitation
(as the symmetry of the mode allows it to be detected in this
experiment~\cite{Diederich-Tunable-Interaction-Excitons-2023}), the energy of which can be expressed as:
\begin{equation}
\small \omega_{-, k=0} = \sqrt{4S\mu_B d^a B_0 \bigg( 1 - \bigg(
\frac{B_{ext}}{B_0}\bigg)^2 \bigg)} \label{eq:magnon}
\end{equation}
where: $S = \tfrac32$ is the spin, $\mu_B$ is the Bohr magneton, $d^a$ is the magnetic anisotropy
along the \textit{a}-axis (shifted so that the anisotropy field
along \textit{b}-axis is zero), $B_{ext}$ is the external magnetic
field, and $B_0$ is the above-mentioned saturation field along the
\textit{c}-axis, which depends on the interlayer spin exchange.

The distinct advantage of CrSBr is that the magnon frequency can be
studied by means of all-optical experiment of time-resolved
reflectivity (TRR)~\cite{bae_exciton-coupled_2022}. In this experiment
the sample is excited non-resonantly by a second-harmonic pulse of
ytterbium femtosecond oscillator (i.e., at 515 nm) and subsequent
near-infrared femtosecond pulse is used to probe the reflectivity at a
given wavelength near the $\Omega_1$ excitonic resonance. By varying
the temporal delay between the pulses we monitor the time-dependent
changes in the reflectivity. As shown in Fig.~\ref{fig:magnony}(a),
such transient reflectivity clearly evidences oscillations that are
unequivocally attributed to the coherent magnon population created by the pump pulse~\cite{bae_exciton-coupled_2022,Diederich-Tunable-Interaction-Excitons-2023}.

The frequency of the observed oscillations varies with the experimental
conditions, in particular with the external magnetic field and the
sample temperature. Importantly, our experiment reveals that the
magnon frequencies vary systematically also as a function of Cl
content. The observed difference is illustrated in
Fig.~\ref{fig:magnony}(b-c), which shows the low-temperature
($T=5\,$K) field-dependence of the magnon spectrum for pristine CrSBr
and for CrSBr$_{0.67}$Cl$_{0.33}$. In both cases (as well as for the
other studied substitution levels), the magnon frequency follows the
predicted $\propto \sqrt{1-(B/B_0)^2}$ dependence [cf.
Eq.~(\ref{eq:magnon})], proving this model as suitable to describe the
magnetic properties of CrSBr$_{1-x}$Cl$_{x}$ alloys for chlorine
substitutions reaching 56\%.

Additionally, the values of the magnetic parameters in Eq.~(\ref{eq:magnon}) systematically vary for the studied samples. In order
to evaluate this change, we fitted the dependence given by Eq.~(\ref{eq:magnon}) to the data series (see [SM]). As the result of
the fit we obtained the values of the two independent intrinsic
magnetic parameters: the \textit{a}-axis magnetic anisotropy $d^a$ and
the saturation field $B_0$. The latter value can be directly compared
with the saturation fields determined earlier using time-integrated PL
and reflectivity spectra, corroborating the trend presented in Fig.~\ref{fig:4}.
The other parameter extracted from the fit, namely the \textit{a}-axis
magnetic anisotropy $d^a$ is a new result delivered by the pump-probe
technique. Additionally, we believe that calculating this parameter
through fit to the data of magnon dispersion in magnetic field grants
much greater precision than calculating through the measurement of
saturation field along all three axes.
Values of both fitted parameters are collected in Fig.~\ref{fig:magnony}(d).

For the magnetic anisotropy $d^a$, we can observe that it decreases significantly in value with the increasing chlorine content and a
considerable deviation from the linear dependence, especially for
contents approaching 56\%. Extrapolating the trend shown by this
parameter, we can observe that the anisotropy along the \textit{a}-axis
reaches the value of 0 near the critical chlorine content of
60-65\%, which falls close to the critical content for phase
transition observed in Ref. \onlinecite{telford_designing_2023}, here described as an A-type AFM to spin glass phase transition.

We will explain this transition more thoroughly by delving into the
structure of the magnetic Hamiltonian. The Cr ions in CrSBr (from the
top-view) form a rectangular lattice, which results in a magnetic
anisotropy favouring the \textit{b}-axis. On the other hand, Cr ions
in CrSCl form a much more isotropic triangular
lattice~\cite{wangPTypeVdWSemiconductor2025}. For such a structural
change to occur, the structural and the in-plane magnetic anisotropy
in the CrSBr$_{1-x}$Cl$_{x}$ alloy have to be lifted with increasing
$x$, a direct evidence of which we see in the fitted $d^a$ parameter.
At last, we see the transition to the in-plane isotropic system at the
critical chlorine substitution, which for our data is around
$x^{Cl}_{crit} \approx 60-65$\%. Over this value,
the Heisenberg interaction between spins has to be supplemented by a
uni-axial anisotropy term (along the \textit{c}-axis),
resulting in the U(1)-symmetric Hamiltonian with a uni-axial
anisotropy (along the \textit{c}-axis):
\begin{equation}
\small \hat H = \sum_{ \langle  ij \rangle } J_{ij} \hat{\mathbf{S}}_i
\cdot \hat{\mathbf{S}}_j + d^c \sum_i (\hat S_i^c)^2.
\end{equation}
The spontaneous breaking of the U(1) symmetry gives rise to a gapless
Goldstone mode, which for $x \ge x_{crit}$ is the magnon mode detected
in the experiment, evidenced with the decrease of the zero-field
frequency to zero as $d^a = 0$.

Fundamentally different information on the system can be gained from the change in the saturation field $B_0$. This parameter depends on the out-of-plane parameters of the material: anisotropy in $c$-axis $d^c$ and the interlayer exchange $J_\perp$. For the antiferromagnetic CrSBr this parameter can be modelled with a dependence:
\begin{equation}\label{eq:B_af}
\small B_0^{AFM} = S \frac{d^c + 2 J_\perp}{\mu_B}.
\end{equation}
In the experiment, we see that as the content of the chlorine increases, the saturation field decreases linearly from around $2.05\,$T to around $1.19\,$T for the maximum chlorine content of 56\%. It is worth noting that this parameter cannot be as easily extrapolated to the limiting content of $100\%$.
As the crystallographic structure switches from the rectangular lattice of CrSBr to the triangular lattice of CrSCl, the interlayer exchange integral changes character from antiferromagnetic to ferromagnetic, as evidenced in Ref.~\cite{wangPTypeVdWSemiconductor2025} and explained by the Goodenough-Kanamori-Anderson rules~\cite{GKA1,GKA2}.
With this sign change, the ground
state without the external magnetic field is ferromagnetically ordered, and the saturation field does not need to compensate
the magnetic order change. This affects the expression for the saturation field $B_0$ from Eq.~\eqref{eq:B_af}, which in the ferromagnetic (FM) case  depends only on the anisotropy parameter:
\begin{equation}
\small B_0^{FM} = S \frac{d^c}{\mu_B}
\end{equation}
Using this expression and a value of $d^c\approx40\,\upmu$eV
\cite{wangPTypeVdWSemiconductor2025}, we would expect that the
saturation field for CrSCl should be: $B_0(x=1) \approx
1.03$\,T. This value was marked in Fig.~\ref{fig:magnony}(d) with a dashed line. By the linear extrapolation of the dependence presented in Fig.~\ref{fig:magnony}(d), this value of the saturation field could be reached around $x \approx 65\%$, marking the threshold for the saturation in the change of $B_0$.

\begin{figure}[H]
    \includegraphics[width = 1\linewidth]{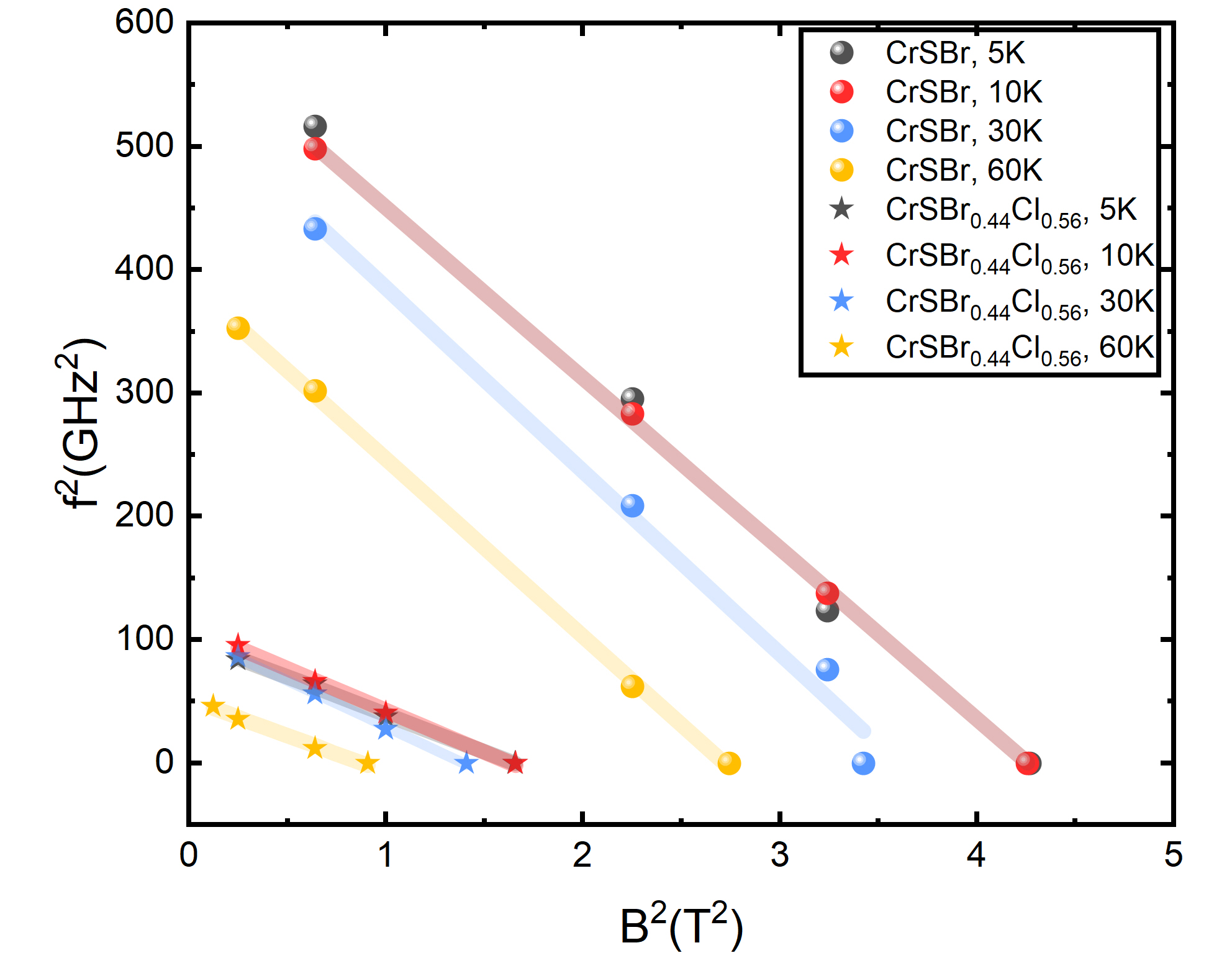}
    \caption{Squared magnon frequency ($f^2$) plotted against the squared magnetic field ($B^2$) measured at temperatures of 5, 10, 30, and 60~K for pristine $\text{CrSBr}$ (circles) and $\text{CrSBr}_{0.44}\text{Cl}_{0.56}$ (stars). Solid lines represent linear fits to the experimental data.}
    \label{fig:temperaturowe}
\end{figure}

It is important to stress that so far all the results and their discussion was based on the experiments conducted at low temperature conditions ($T=5\,$K). It triggers the question of their applicability at elevated temperatures. In order to address this issue we have performed an additional series of measurements. For practical reasons we focused on two distinct samples: CrSBr$_{1-x}$Cl$_x$ with $x=56\%$ and the pristine CrSBr (i.e., $x=0$), for which we performed pump-probe experiments at several magnetic fields for different temperatures. The extracted frequencies of the magnon oscillations are presented in Fig.~\ref{fig:temperaturowe} and plotted in coordinates $f^2$ vs. $B^2$. According to Eq.~\eqref{eq:magnon}, in such coordinate system the expected $f^2(B^2)$ dependence is a straight line with a slope that is directly proportional to $d^a/B_0$. The presented data reveal that (1) this slope for a given sample is independent of the temperature while (2) the slope is clearly different for different samples. Combined, it unequivocally proves that the distinction between the samples with different Cl content persists also at elevated temperatures.

%

This qualitative difference between the constant slope and the varying intercept of $f^2(B^2)$ dependence in Fig. \ref{fig:temperaturowe} can be understood as the effect of the incoherent (thermal) magnons populating the sample at finite temperatures.
This population at low temperatures is proportional to the temperature ($n\propto T/T_\mathrm{corr}$ with $T_\mathrm{corr} \approx 180\,$K)\cite{dirnberger2023magneto} and introduces quantum fluctuations to the system.
These fluctuations in turn decrease the magnetization of the layers with the dependence: $(M_s/M_0)^2 \approx 1 - 2\beta T/T_\mathrm{corr}$ \cite{dirnberger2023magneto}.
On the microscopic level, it can be understood as the quantum fluctuations $\delta S$ entering the spin $S$, which appears in both $f$ and $B_0$.
Now, we can see that the temperature dependence will vanish in the slope parameter, proportional to $S d^a/B_0$, while it will introduce the linear dependence on temperature into the intercept, being a function of $S d^a B_0$.
This analysis assumes no temperature dependence of the other microscopic magnetic parameters ($d^a, d^c, J_\perp$), which in a fully physical picture should itself be dependent on the temperature.
The qualitative agreement between above mean-field-like considerations and the experimental data in Fig. \ref{fig:temperaturowe} is a strong suggestion that such a simple description is valid for the studied samples of different Cl content at least in the temperature range up to 60 K.
\section{Conclusions} \label{sec:conclusions}
Various experimental techniques employed in our investigation paint a consistent picture:
partial Br$\to$ Cl substitution in CrSBr \emph{continuously} affects the crucial magneto-optical properties of the crystals. This observation is confirmed over the full range of iinvestigated samples, i.e., up to 56\% substitution. 

The tunability obtained using this approach is significant, reaching reduction of the saturation point to nearly half of its reference value for pristine CrSBr. Similarly, the zero-field magnon frequency decreases from 24 GHz down to 10 GHz. The variation of these parameters is nearly linear with the chlorine content, hinting the description within virtual crystal approximation. 

\par
Importantly, for the whole scope of studied samples the reported variation of the discussed parameters did not hinder the coupling between magnetic and excitonic properties, which is a hallmark feature of CrSBr. As we show, all observed excitonic transitions experience a large energy shift which reflects  modification of the relative magnetization of adjacent layers. The strength of this coupling remains llargely constant with the chlorine content, even though the absolute energy of all observed excitons is shown to be sensitive to the Cl concentration.

Overall, our findings provide a quantitative description of how chemical modification influences the magneto-optical response of CrSBr and offer a solid foundation for development of optically active heterostructures of tailored magnetic or magnonic landscape.

\section*{Acknowledgements} 
\label{sec:acknowledgements}
This work was supported by the Polish National Science Centre, Poland under decisions DEC-2020/38/E/ST3/00364 and 2024/55/B/ST3/03144. We also acknowledge project LUAUS25268 from Ministry of Education Youth and Sports (MEYS) and by the Advanced Multiscale Materials for Key Enabling Technologies project, supported by the Ministry of Education, Youth, and Sports of the Czech Republic. Project No. CZ.02.01.01/00/22$\_$008/0004558, Co-funded by the European Union. 
The authors gratefully acknowledge Marta Bilska for her assistance with the EDX measurements and Bartosz Furtak for his assistance with the Raman spectroscopy measurements.


\bibliography{sections/crsbr}

@article{crsbr-magnetoresistance-2020,
author = {Telford, Evan J. and Dismukes, Avalon H. and Lee, Kihong and Cheng, Minghao and Wieteska, Andrew and Bartholomew, Amymarie K. and Chen, Yu-Sheng and Xu, Xiaodong and Pasupathy, Abhay N. and Zhu, Xiaoyang and Dean, Cory R. and Roy, Xavier},
title = {Layered Antiferromagnetism Induces Large Negative Magnetoresistance in the van der Waals Semiconductor CrSBr},
journal = {Advanced Materials},
volume = {32},
number = {37},
pages = {2003240},
doi = {https://doi.org/10.1002/adma.202003240},
year = {2020}
}

@article{crsbr-wilson-2021,
  title = {Interlayer Electronic Coupling on Demand in a {{2D}} Magnetic Semiconductor},
  author = {Wilson, Nathan P. and Lee, Kihong and Cenker, John and Xie, Kaichen and Dismukes, Avalon H. and Telford, Evan J. and Fonseca, Jordan and Sivakumar, Shivesh and Dean, Cory and Cao, Ting and Roy, Xavier and Xu, Xiaodong and Zhu, Xiaoyang},
  year = 2021,
  journal = {Nature Materials},
  volume = {20},
  number = {12},
  pages = {1657--1662},
  publisher = {Nature Research},
  issn = {1476-1122},
  doi = {10.1038/s41563-021-01070-8},
 year = {2021}
}

@article{cenker_reversible_2022,
  title = {Reversible Strain-Induced Magnetic Phase Transition in a van Der {{Waals}} Magnet},
  author = {Cenker, John and Sivakumar, Shivesh and Xie, Kaichen and Miller, Aaron and Thijssen, Pearl and Liu, Zhaoyu and Dismukes, Avalon and Fonseca, Jordan and Anderson, Eric and Zhu, Xiaoyang and Roy, Xavier and Xiao, Di and Chu, Jiun-Haw and Cao, Ting and Xu, Xiaodong},
  year = {2022},
  month = mar,
  journal = {Nature Nanotechnology},
  volume = {17},
  number = {3},
  pages = {256--261},
  publisher = {Nature Publishing Group},
  issn = {1748-3395},
  doi = {10.1038/s41565-021-01052-6},
  urldate = {2025-10-31},
  copyright = {2022 The Author(s), under exclusive licence to Springer Nature Limited},
  langid = {english}
}

@article{bae_exciton-coupled_2022,
  title = {Exciton-Coupled Coherent Magnons in a {{2D}} Semiconductor},
  author = {Bae, Youn Jue and Wang, Jue and Scheie, Allen and Xu, Junwen and Chica, Daniel G. and Diederich, Geoffrey M. and Cenker, John and Ziebel, Michael E. and Bai, Yusong and Ren, Haowen and Dean, Cory R. and Delor, Milan and Xu, Xiaodong and Roy, Xavier and Kent, Andrew D. and Zhu, Xiaoyang},
  year = {2022},
  month = sep,
  journal = {Nature},
  volume = {609},
  number = {7926},
  pages = {282--286},
  publisher = {Nature Publishing Group},
  issn = {1476-4687},
  doi = {10.1038/s41586-022-05024-1},
  urldate = {2025-10-31},
  copyright = {2022 The Author(s), under exclusive licence to Springer Nature Limited},
  langid = {english}
}

@article{goser_magnetic_1990,
	title = {Magnetic properties of {CrSBr}},
	volume = {92},
	issn = {0304-8853},
	url = {https://www.sciencedirect.com/science/article/pii/030488539090689N},
	doi = {10.1016/0304-8853(90)90689-N},
	number = {1},
	urldate = {2025-10-31},
	journal = {Journal of Magnetism and Magnetic Materials},
	author = {Göser, O. and Paul, W. and Kahle, H. G.},
	month = nov,
	year = {1990},
	pages = {129--136},

}

@article{lee_magnetic_2021,
  title = {Magnetic {{Order}} and {{Symmetry}} in the {{2D Semiconductor CrSBr}}},
  author = {Lee, Kihong and Dismukes, Avalon H. and Telford, Evan J. and Wiscons, Ren A. and Wang, Jue and Xu, Xiaodong and Nuckolls, Colin and Dean, Cory R. and Roy, Xavier and Zhu, Xiaoyang},
  year = {2021},
  month = apr,
  journal = {Nano Letters},
  volume = {21},
  number = {8},
  pages = {3511--3517},
  publisher = {American Chemical Society},
  issn = {1530-6984},
  doi = {10.1021/acs.nanolett.1c00219},
  urldate = {2025-10-31}
}

@article{heissenbuttel-quadratic-response-crsbr,
  title = {Quadratic optical response to a magnetic field in the layered magnet CrSBr},
  author = {Hei\ss{}enb\"uttel, Marie-Christin and Piel, Pierre-Maurice and Klein, Julian and Deilmann, Thorsten and Wurstbauer, Ursula and Rohlfing, Michael},
  journal = {Phys. Rev. B},
  volume = {111},
  issue = {7},
  pages = {075107},
  numpages = {6},
  year = {2025},
  month = {Feb},
  publisher = {American Physical Society},
  doi = {10.1103/PhysRevB.111.075107},
  url = {https://link.aps.org/doi/10.1103/PhysRevB.111.075107}
}

@article{telford_designing_2023,
  title = {Designing {{Magnetic Properties}} in {{CrSBr}} through {{Hydrostatic Pressure}} and {{Ligand Substitution}}},
  author = {Telford, Evan J. and Chica, Daniel G. and Ziebel, Michael E. and Xie, Kaichen and Manganaro, Nicholas S. and Huang, Chun-Ying and Cox, Jordan and Dismukes, Avalon H. and Zhu, Xiaoyang and Walsh, James P. S. and Cao, Ting and Dean, Cory R. and Roy, Xavier},
  year = {2023},
  journal = {Advanced Physics Research},
  volume = {2},
  number = {11},
  pages = {2300036},
  issn = {2751-1200},
  doi = {10.1002/apxr.202300036},
  urldate = {2025-10-28},
  copyright = {{\copyright} 2023 The Authors. Advanced Physics Research published by Wiley-VCH GmbH},
  langid = {english}
}

@article{zhang-crocl,
  title = {Magnetism and {{Optical Anisotropy}} in van Der {{Waals Antiferromagnetic Insulator CrOCl}}},
  author = {Zhang, Tianle and Wang, Yimeng and Li, Hexuan and Zhong, Fang and Shi, Jia and Wu, Minghui and Sun, Zhaoyang and Shen, Wanfu and Wei, Bin and Hu, Weida and Liu, Xinfeng and Huang, Li and Hu, Chunguang and Wang, Zhongchang and Jiang, Chengbao and Yang, Shengxue and Zhang, Qing-ming and Qu, Zhe},
  year = {2019},
  month = oct,
  journal = {ACS Nano},
  volume = {13},
  number = {10},
  pages = {11353--11362},
  publisher = {American Chemical Society},
  issn = {1936-0851},
  doi = {10.1021/acsnano.9b04726},
  urldate = {2025-10-28}
}

@article{han-crsebr,
  title = {Enhanced Ferromagnetism, Perpendicular Magnetic Anisotropy and High {{Curie}} Temperature in the van Der {{Waals}} Semiconductor {{CrSeBr}} through Strain and Doping},
  author = {Han, Ruilin and Xue, Xiaomin and Li, Peng},
  year = {2024},
  month = apr,
  journal = {Physical Chemistry Chemical Physics},
  volume = {26},
  number = {15},
  pages = {12219--12230},
  publisher = {The Royal Society of Chemistry},
  issn = {1463-9084},
  doi = {10.1039/D4CP00855C},
  urldate = {2025-10-28},
  langid = {english}
}

@article{Diederich-Tunable-Interaction-Excitons-2023,
  title = {Tunable Interaction between Excitons and Hybridized Magnons in a Layered Semiconductor},
  author = {Diederich, Geoffrey M. and Cenker, John and Ren, Yafei and Fonseca, Jordan and Chica, Daniel G. and Bae, Youn Jue and Zhu, Xiaoyang and Roy, Xavier and Cao, Ting and Xiao, Di and Xu, Xiaodong},
  year = {2023},
  month = jan,
  journal = {Nature Nanotechnology},
  volume = {18},
  number = {1},
  pages = {23--28},
  publisher = {Nature Publishing Group},
  issn = {1748-3395},
  doi = {10.1038/s41565-022-01259-1},
  urldate = {2026-01-21},
  copyright = {2022 The Author(s), under exclusive licence to Springer Nature Limited},
  langid = {english}
}

@article{Smiertka-Distinct-Magnetooptical-Response2026,
  title = {Distinct Magneto-Optical Response of {{Frenkel}} and {{Wannier}} Excitons in {{CrSBr}}},
  author = {{\'S}miertka, Maciej and Ryga{\l}a, Micha{\l} and Posmyk, Katarzyna and Peksa, Paulina and Dyksik, Mateusz and Pashov, Dimitar and Mosina, Kseniia and Sofer, Zden{\v e}k and {van Schilfgaarde}, Mark and Dirnberger, Florian and Baranowski, Micha{\l} and Acharya, Swagata and Plochocka, Paulina},
  year = {2026},
  month = jan,
  journal = {Nature Communications},
  publisher = {Nature Publishing Group},
  issn = {2041-1723},
  doi = {10.1038/s41467-026-68482-5},
  urldate = {2026-01-21},
  copyright = {2026 The Author(s)},
  langid = {english}
}

@article{Datta-Magnon-mediated-ExcitonExciton-2025,
  title = {Magnon-Mediated Exciton--Exciton Interaction in a van Der {{Waals}} Antiferromagnet},
  author = {Datta, Biswajit and Adak, Pratap Chandra and Yu, Sichao and Valiyaparambil Dharmapalan, Agneya and Hall, Siedah J. and Vakulenko, Anton and Komissarenko, Filipp and Kurganov, Egor and Quan, Jiamin and Wang, Wei and Mosina, Kseniia and Sofer, Zden{\v e}k and Pashov, Dimitar and {van Schilfgaarde}, Mark and Acharya, Swagata and Kamra, Akashdeep and Sfeir, Matthew Y. and Al{\`u}, Andrea and Khanikaev, Alexander B. and Menon, Vinod M.},
  year = {2025},
  month = jul,
  journal = {Nature Materials},
  volume = {24},
  number = {7},
  pages = {1027--1033},
  publisher = {Nature Publishing Group},
  issn = {1476-4660},
  doi = {10.1038/s41563-025-02183-0},
  urldate = {2026-01-21},
  copyright = {2025 The Author(s), under exclusive licence to Springer Nature Limited},
  langid = {english}
}

@article{wang-Exciton-polaritons2023,
  title = {Magnetically-Dressed {{CrSBr}} Exciton-Polaritons in Ultrastrong Coupling Regime},
  author = {Wang, Tingting and Zhang, Dingyang and Yang, Shiqi and Lin, Zhongchong and Chen, Quan and Yang, Jinbo and Gong, Qihuang and Chen, Zuxin and Ye, Yu and Liu, Wenjing},
  year = {2023},
  month = sep,
  journal = {Nature Communications},
  volume = {14},
  number = {1},
  pages = {5966},
  publisher = {Nature Publishing Group},
  issn = {2041-1723},
  doi = {10.1038/s41467-023-41688-7},
  urldate = {2024-01-04},
  copyright = {2023 The Author(s)},
  langid = {english}
}

@article{Rafał-arxiv,
  title={Colossal magneto-excitonic effects in 2d van der Waals magnetic semiconductor CrSBr},
  author={R. Komar and A. Łopion and M. Goryca and M. Rybak and T. Woźniak and K. Mosina and A. Söll and Z. Sofer and W. Pacuski and C. Faugeras and M. Birowska and P. Kossacki and T. Kazimierczuk},
  journal={arXiv preprint arXiv:2409.00187},
  year={2024}
}

@article{wangPTypeVdWSemiconductor2025,
  title = {P-{{Type vdW Semiconductor CrSCl Featuring Multipolarity Coexistence}}},
  author = {Wang, Yong and Yang, Dingyi and Wang, Shaopeng and Xu, Wei and Zhang, Yu and Johal, Tarnjit Kaur and Xu, Yongjie and Wang, Yongmei and Zhang, Yihan and Zhang, Yin and Bai, Wubin and Wu, Yizhang and Hao, Yue},
  year = {2025},
  month = feb,
  journal = {ACS Materials Letters},
  volume = {7},
  number = {2},
  pages = {636--645},
  publisher = {American Chemical Society},
  doi = {10.1021/acsmaterialslett.4c02378},
  urldate = {2026-04-29},
}

@article{badolaVanWaalsCrSBr2026,
  title = {Van Der {{Waals CrSBr Alloys}} with {{Tunable Magnetic}} and {{Optical Properties}}},
  author = {Badola, Shalini and Pawbake, Amit and Wu, Bing and S{\"o}ll, Aljoscha and Sofer, Zdenek and Heid, Rolf and Faugeras, Clement},
  year = {2026},
  month = jan,
  journal = {Nano Letters},
  volume = {26},
  number = {2},
  pages = {773--779},
  publisher = {American Chemical Society},
  issn = {1530-6984},
  doi = {10.1021/acs.nanolett.5c05146},
  urldate = {2026-04-29}
}

@article{GKA1,
  title = {Theory of the Role of Covalence in the Perovskite-Type Manganites $[\mathrm{La}, M(\mathrm{II})]\mathrm{Mn}{\mathrm{O}}_{3}$},
  author = {Goodenough, John B.},
  journal = {Phys. Rev.},
  volume = {100},
  issue = {2},
  pages = {564--573},
  numpages = {0},
  year = {1955},
  month = {Oct},
  publisher = {American Physical Society},
  doi = {10.1103/PhysRev.100.564},
  url = {https://link.aps.org/doi/10.1103/PhysRev.100.564}
}

@article{GKA2,
title = {Superexchange interaction and symmetry properties of electron orbitals},
journal = {Journal of Physics and Chemistry of Solids},
volume = {10},
number = {2},
pages = {87-98},
year = {1959},
issn = {0022-3697},
doi = {https://doi.org/10.1016/0022-3697(59)90061-7},
url = {https://www.sciencedirect.com/science/article/pii/0022369759900617},
author = {Junjiro Kanamori}
}

@article{dirnberger2023magneto,
  title={Magneto-optics in a van der Waals magnet tuned by self-hybridized polaritons},
  author={Dirnberger, Florian and Quan, Jiamin and Bushati, Rezlind and Diederich, Geoffrey M and Florian, Matthias and Klein, Julian and Mosina, Kseniia and Sofer, Zdenek and Xu, Xiaodong and Kamra, Akashdeep and others},
  journal={Nature},
  volume={620},
  number={7974},
  pages={533--537},
  year={2023},
  publisher={Nature Publishing Group UK London}
}

\end{document}